\documentclass[twocolumn]{aa}
\usepackage{psfig}
\usepackage{txfonts}
\def\deg{$^{\circ}$}
\def\mvs{$\overline{v\sin i}$}
\def\vsini{$v$\,$\sin i$}
\def\vrad{$v_{\rm rad}$}
\def\kms{km/s}
\def\texp{$<T>$}
\def\tmax{$T_{\rm max}$}
\def\t90{$T_{\rm min90}$}
\def\msol{M$_{\odot}$}
\def\rsol{R$_{\odot}$}
\begin{document}

\title{Spectroscopic rotation velocities of L dwarfs from VLT/UVES and their
comparison with periods from photometric monitoring
\thanks{Based on observations obtained at the VLT, European Southern Observatory, Chile, program 65.L-0199}}
\titlerunning{Rotation velocities of L dwarfs}

\author{C.A.L.\ Bailer-Jones\inst{1}\fnmsep\inst{2}\fnmsep\thanks{Emmy Noether Fellow of the Deutsche Forschungsgemeinschaft}}

\institute{Max-Planck-Institut f\"ur Astronomie, K\"onigstuhl 17, 69117
Heidelberg, Germany
\and Carnegie Mellon University, Department of Physics, 5000
Forbes Ave., Pittsburgh, PA 15213, USA\\
\email{calj@mpia-hd.mpg.de}
}

\date{Submitted 3 January 2004; Revised 19 February 2004}

\abstract{The variability and rotation of ultra cool dwarfs (UCDs)
provide important information on the atmospheres and evolution of
these very low mass stars and brown dwarfs.  As part of an ongoing
program to investigate this, the projected rotation periods, \vsini,
derived from high resolution VLT/UVES spectroscopy via cross
correlation are presented for 16 field UCDs (M9V--L7.5V). This doubles
the number of L dwarfs for which \vsini\ has been measured.  All
targets are found to have \vsini\ between 10 and 40\,\kms\ confirming
that L dwarfs are rapid rotators.  Radial velocities have also been
measured to a precision of 1--2\,\kms.  From the random distribution
of the rotation axes, $i$, and theoretically predicted radii,
one-sided confidence intervals are placed on the rotation periods of
individual objects. These are compared with published period data
obtained from photometric monitoring programs.  From this, the period
of 31\,hrs for the L0.5 dwarf \object{2M0746+2000} published by Gelino
et al.\ (\cite{gelino02}) may be ruled out as the rotation period.
The period of 11.2\,$\pm$\,0.8\,hrs for the L1.5 dwarf
\object{2M1145+2317} obtained by Bailer-Jones \& Mundt (\cite{bjm01})
is consistent with the present \vsini\ results so is plausibly the
true rotation period.  The inclination of the rotation axis is
constrained to be $i$=62\deg--\,90\deg\ with an expectation value of
76\deg.  Alternatively the data set a lower limit on the radius of
0.1\rsol, which is within the range of radii predicted by models for
brown dwarfs older than 0.5\,Gyr.  Similarly, the period of
2.7\,$\pm$\,0.1\,hrs detected by the same authors for
\object{2M1334+1940} is also confirmed as the likely rotation period;
the inclination is $i$=27\deg--\,44\deg ($<i>=34$\deg).  Where no
variability or period was detected by the monitoring programs the
likely reason is low contrast modulating surface features.  However,
in three cases variability but no period was detected, even though the
likely rotation period range inferred from \vsini\ lies within the
timescale to which the monitoring was sensitive.  This reinforces the
`masking hypothesis' of Bailer-Jones \& Mundt (\cite{bjm01}), the idea
that the evolution of photospheric features on timescales shorter than
the rotation period obscure the regular modulation of the light curve.
As has been previously discussed, a likely candidate for such features
is inhomogeneous dust clouds.


}

\maketitle

\section{Introduction}

Understanding the nature of low mass stars and brown dwarfs requires a
detailed characterization of their observable
atmospheres. Specifically, the determination of luminosities, chemical
compositions and ages ultimately depends on adequate knowledge of
the radiative and convective transport mechanisms.  While our
understanding of these mechanisms is reasonable (if incomplete) for
some types of stars, it is comparatively poor for ultra cool dwarfs
(UCDs), late type M, L and T dwarfs. This is not least because at the
low effective temperatures involved ($<$3000\,K),
solid dust particles form in significant numbers and varieties,
and these have a major impact on the structure of the atmosphere and
hence on the observable properties of the star (e.g.\ Allard et al.\
\cite{allard01}; Burrows \& Sharp \cite{burrows99}). In particular,
global dust structures (`clouds') may play a significant role.

 Significant information on UCDs can be
obtained from their observable temporal changes,  and
several dedicated monitoring programs have now uncovered good evidence
for optical and infrared variability (Bailer-Jones \& Mundt
\cite{bjm99}; Terndrup et al.\ \cite{terndrup99}; Tinney \& Tolley
\cite{tinney99}; Nakajima et al.\ \cite{nakajima00}; Bailer-Jones \&
Mundt \cite{bjm01}; Martin et al.\ \cite{martin01}; Bailer-Jones
\cite{bj02}; Burgasser et al.\ \cite{burgasser02}; Clarke et al.\
\cite{clarke02a}, \cite{clarke02b}; Gelino \cite{gelino02}; Gelino et
al.\ \cite{gelinoetal02}; Bailer-Jones \& Lamm \cite{bjl}; Clarke et
al.\ \cite{clarke03}; Enoch et al.\ \cite{enoch03}; Joergens et al.\
\cite{joergens03}; Koen \cite{koen03}; Zapatero Osorio et al.\
\cite{zapatero03}) and H$\alpha$ variability (Hall \cite{hall02};
Mochnacki et al.\ \cite{mochnacki02}; Liebert et al.\
\cite{liebert03}).  Of the 80 or so UCDs monitored by these groups,
about 30 show evidence for photometric variability. While this figure
depends on what one takes as sufficient evidence for variability, at
least half are convincingly variable on the scale of a few tens of
millimagnitudes. There are indications that this photometric
variability is not caused by a simple rotational modulation of a
non-uniform photosphere. One candidate is an inhomogeneous dust layer
evolving under the influence of convection on time scales shorter than
the rotation period (see the introduction to Bailer-Jones \cite{bj02}
for a discussion).  Understanding this phenomenon is important, not
only for what it tells us about atmospheric process under these
interesting physical conditions, but also for the implications it has
concerning conclusions drawn from single epoch observations of UCDs.

In this article I address the specific issue of how the photometric
variability and period data relate to UCD
rotation, as measured by line broadening from high resolution
spectroscopy. This provides the {\it projected} rotation speed, \vsini,
so using simple statistical arguments and theoretical predictions for
UCD radii, constraints can be placed on likely rotation periods.
These (plus other published \vsini\ data) are compared with
the variability data to confirm or refute possible rotation periods
and investigate the `masking hypothesis'. 

Previously published work has established that UCDs are fast rotators:
of the 17 hitherto observed L dwarfs, 16 have \vsini\ in the range
10--40\,\kms\ and one (Kelu-1) has 60\,\kms\ (Basri et al.\
\cite{basri00}; Mohanty \& Basri \cite{mohanty03}; Schweitzer et al.\
\cite{schweitzer01}).  In the present work, \vsini\ values are
determined for 16 UCDs (15 L dwarfs and 1 M dwarf), 14 of which have
no previously published measurement.  Moreover, conservative ranges
for \vsini\ are established which reflect the dominant uncertainties
in measuring \vsini\ by cross correlation. For the majority of UCDs,
the complete \vsini\ range lies between 10 and 30\,\kms, and there is
no measurement below about 10\,\kms, thus confirming the rapid
rotation of UCDs.  For a discussion of UCD rotation and its relation
to chromospheric activity and possible dynamo mechanisms, the reader
is referred to Mohanty \& Basri (\cite{mohanty03}).

\section{Data acquisition and reduction}

A target list of UCDs ranging in spectral type from M9 to L8 was
assembled from the published UCDs available at the time
of the observations (April 2000). They were selected preferentially
for brightness and for good observability from the VLT and to cover a
range of spectral types. They were not selected based on any known
\vsini\ values (none were available at the time), although some
L dwarfs monitored by Bailer-Jones \& Mundt (\cite{bjm99}, \cite{bjm01}) were
specifically included.

\subsection{Instrumentation and observations}

High resolution spectroscopy was obtained with the UVES echelle
spectrograph mounted on the UT2 8.2m VLT telescope at Cerro Paranal,
Chile, on 26--28 April 2000.  The red arm of this instrument was used
with the CD4-prot cross dispersion grating and a slit width of 1$''$.
This provides the wavelength range 6440--10\,250\,\AA\ (orders 94--60)
at a resolution of 38\,700, yielding a FWHM of 7.8\,\kms.  The
detector is a mosaic of two CCDs with pixels binned to provide a
sampling of 2.4\,\kms/pix in the dispersion direction and 0.35$''$/pix
in the spatial direction. The slit length was 10$''$  and was oriented
vertically (i.e.\ parallel to the direction of maximum atmospheric
dispersion). 

Over the course of two and a half nights, spectra were obtained of the
16 UCDs listed in Table~\ref{vsini_table}.  Exposure times ranged from
1$\times$15\,min for the brightest targets to 3$\times$50\,min for the
faintest targets.  The seeing was typically 0.8$''$ (0.5$''$--1.0$''$)
at airmasses of up to 2.0.  Spectra were also obtained several times with the same
instrument settings for \vsini\ and \vrad\ standards,  namely the bright
M dwarfs Gl402, Gl406 and Gl876.  The spectrograph is
mounted on the Nasmyth platform so undergoes limited movement
during the course of the observations.  The data reduction procedure
verified the stability of the instrument. While wavelength
calibration spectra and flat fields were taken at regular
intervals each night, it was found that a single set of calibration
frames for each night was sufficient.

\subsection{Data reduction and spectral extraction}\label{datared}

The objective of the data reduction is to produce wavelength
calibrated one-dimensional spectra for each echelle order, free of
relevant instrumental and telluric signatures as preparation for
cross correlation.  The basic data
reduction was carried out using the IRAF\footnote{the Image Reduction
and Analysis Facility, provided by the National Optical Astronomy
Observatories (NOAO).} package, and consists of the removal of scattered
light, flat fielding, spectral extraction and wavelength calibration.

The two-dimensional data format (i.e.\ cross dispersed spectra) will
be referred to as an ``image''.  Scattered light in the spectrograph
contributes a two-dimensional additive pattern to the dispersed light
and generally needs to be removed from both science frames and flats.
This was done by tracing the orders and defining the inter-order
regions. A two-dimensional fit is made to these regions and this fit
subtracted from all images.  As dark frames, biases and overscan
regions showed no features, a further explicit subtraction of these
was not necessary.

Flat fielding is performed to remove small scale detector variations
and fringing.\footnote{In broad band direct imaging, fringing
(monochromatic interference in the CCD) is generated by night sky line
emission sources. As discussed in Bailer-Jones \& Mundt (\cite{bjm01}) this
produces an additive pattern over the whole CCD which must be
subtracted to perform correct photometry (and not divided, as is often
thought, because it does not modulate the star light).  With high
resolution spectroscopy the situation is different.  Here the grating
itself creates an independent narrow band source at every point along
the dispersion direction which in turn generates fringing at each
point in the CCD.  All incident sources therefore create the same
fringe pattern at a given wavelength, so the fringes must be divided
out using a flat field.}  A sequence of individual flat field images
is combined.  This master flat was then rectified using the {\sc
apflatten} task, which makes fits to each order to remove the
pseudo-continuum and grating blaze function.  The result is a
two-dimensional flat field preserving only high frequency variations
which are then removed from the science frames by division.  The
global variations (pseudo-continuum, blaze function etc.) are removed
later immediately prior to cross correlation.

The spectra are extracted and the sky subtracted using the {\sc apsum}
package. A one-dimensional optimal extraction was performed with pixel
cleaning based on expected noise statistics.  Because many of the L dwarfs
are faint, extraction apertures are not traced: they are transferred
from the order definition image and simply recentered and resized to
accommodate both the position of the sources in the slit and the seeing.

A wavelength calibration was established using the internal ThAr
lamp. Lines were iteratively identified and a fit produced using the
{\sc ecreidentify} package.  The fit was a
two-dimensional Chebyshev polynomial based on the spectral order
numbers (using a third order polynomial along the dispersion axis and
second order one perpendicular to it).  The RMS of the
fit was 0.01 pixels, or 25\,m/s. This was applied to the science
spectra using the {\sc dispcor} package. Changes in temperature or
pressure or seismic disturbances can alter the zero point of this
calibration during the course of a night. The calibration was
therefore checked in all science images against the positions of the
OH emission lines from the Earth's atmosphere. These demonstrated that
the zero point shift of the instrumental calibration was below the RMS
of the fit.   Of course, additional zero point shifts
occur in the applied wavelength scale if a star is not perfectly
centered in slit, and could be as large as the projected slit size. In
practice, though, autoguiding keeps the star centered in the slit to
within about a tenth of the FWHM, giving an upper limit on the
systematic zero point shift (and hence systematic radial velocity
error) of about 1\,\kms.  This does not effect the rotational
velocities.  

\begin{figure}
\centerline{
\psfig{figure=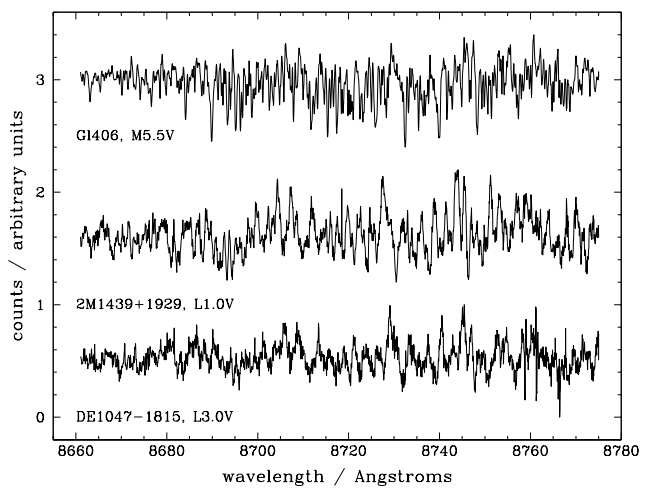,angle=270,width=0.50\textwidth}
}
\caption{Spectra (order 70) of an M dwarf \vsini\ and \vrad\ standard
(top), an L1.0 dwarf (also the L dwarf `standard', middle), and an
L3.0 dwarf (bottom). The spectra have been continuum subtracted,
scaled to a common vertical range and offset on the vertical axis.
The exposure times for these single spectra were 3\,min, 15\,min and
40\,min respectively, which achieve mean SNR per resolution element
around the centre of the order of 150, 15 and 11 respectively. }
\label{specplot}
\end{figure}

 Candidate wavelength regions for the cross
correlation were selected on the grounds of being relatively free of
telluric absorption, as identified via inspection of stellar spectra
with few intrinsic features (a rapidly rotating B9V, a B5V and an
A0III) acquired for this purpose.  As a result, just those orders (or
parts of orders) listed in Table~\ref{orders} are retained for all
targets and templates. Regions containing strong telluric emission
lines (which, on account of noise, are only poorly removed by sky
subtraction) were likewise removed.  While performing the cross
correlations on individual objects, all orders were visually inspected
and any obvious emission lines, cosmic rays or bad pixels were masked. Not all
orders in Table~\ref{orders} were used to form the final values of
\vsini\ or \vrad: orders were selected for each star separately based
on the quality of the cross correlation function (see below).  Example
spectra for three objects are shown in Fig.~\ref{specplot}. The
signal-to-noise ratio (SNR) varies for different orders and different
targets, but in order 70 it is 8--25 per extracted pixel (0.05\,\AA; Cf.\
FWHM of a resolution element of 0.23\,\AA\ in order 70). Orders with SNR
lower than a few were disregarded; for this reason the bluest orders
were often not used.  

\begin{table}
\begin{center}
\caption{Candidate orders or partial orders retained for the cross
correlation.  In individual cases, some orders produce poor
correlation functions in either the cross correlation of the template
against the target or, for \vsini, the template against the spun up
template (calibration), in which case the order is not used.  Orders
90--94 were not used with the L dwarf rotation template, and
only orders 60--72 were use in determining \vrad, in both cases for
SNR reasons.}
\label{orders}
\begin{tabular}{lrr}
\hline
order & $\lambda_{\rm min.}$/\AA & $\lambda_{\rm max.}$/\AA \\
\hline
60 & 10095.0 & 10250.0 \\
61 &  9931.0 & 10080.0 \\
62 &  9850.0 &  9917.0 \\
69 &  8786.0 &  8903.0 \\
70 &  8661.0 &  8775.0 \\
71 &  8540.0 &  8651.0 \\
72 &  8419.0 &  8530.0 \\
75 &  8087.0 &  8130.0 \\
76 &  7979.5 &  8078.0 \\
77 &  7877.0 &  7973.0 \\
78 &  7777.0 &  7869.0 \\
79 &  7699.0 &  7770.0 \\
81 &  7490.0 &  7577.0 \\
82 &  7399.0 &  7484.0 \\
85 &  7139.0 &  7165.0 \\
86 &  7054.0 &  7133.0 \\
89 &  6820.0 &  6866.0 \\
90 &  6745.0 &  6815.0 \\
91 &  6670.0 &  6739.0 \\
92 &  6598.0 &  6665.0 \\
94 &  6460.0 &  6523.0 \\
\hline
\end{tabular}
\end{center}
\end{table}

\section{Radial velocities}

Radial velocities are determined via cross correlation of the UCD
spectra with the spectrum of an observed star of known radial
velocity. This was done in two stages.  The initial template used was
Gl406, an M4V star with \vrad\ reported as 19.48\,$\pm$\,0.03\,\kms\
by Nidever et al.\ (\cite{nidever02}) and 19.18\,$\pm$\,0.11\,\kms\ by
White \& Basri (\cite{white03}).  I adopt the former. The cross
correlation was performed using the IRAF package {\sc fxcor}, with the
centre of the cross correlation peak calculated using the `center1d'
option. The pseudo-continuum of each order is rectified (fit and
divided out) using a fifth order cubic spline.

The cross correlation function was sometime quite noisy, in particular
for the bluer spectral orders and for late L spectral types, resulting
in ambiguous peak centering or even identification.  For this reason,
the bright L1.0 dwarf 2M1439+1929 was adopted as a `secondary' radial
velocity standard, and only the orders in the `red' CCD (orders 60,
61, 62, 69, 70, 71 and 72 in Table~\ref{orders}) are used. The radial
velocity of 2M1439+1929 is determined to be -26.3\,$\pm$\,0.50\,\kms\
from cross correlation against Gl406. (The bright L0.5 dwarf
2M0746+2000 could also have been used for this purpose, but was not on
account of its binarity; section~\ref{assessment}).  The radial
velocities for all other UCDs are determined by cross correlation
against 2M1439+1929 and are listed in Table~\ref{vsini_table}: the
value given is a weighted mean of the results in the different
orders. The uncertainty is the standard deviation in this mean
(including the uncertainty in the template velocity), apart from for a
few cases in which the peak centering uncertainty was larger, in which
case this is reported instead.

Three of the results are dubious (marked with a ``:'') because of
discrepancies between orders and/or significant centering
uncertainties due to highly rotationally broadened spectral lines. The
latter is particularly applicable for DE1159+0057 and DE1431-1953.
Two of the targets have \vrad\ determinations in the literature.
Basri et al.\ (\cite{basri00}) obtained \vrad\,=\,29\,$\pm$\,1\,\kms\
for 2M1439+1929, which differs from my result by 2.5$\sigma$.  Reid et
al.\ (\cite{reid02}), on the other hand, found
\vrad\,=\,54.1\,$\pm$\,0.80\,\kms\ for 2M0746+2000, in excellent
agreement with my result. Recall that my results could have an
additional systematic error of up to 1\,\kms\ (section~\ref{datared}).

I should note that Basri et al.\ (\cite{basri00}) have observed Gl406
to have a significantly different radial velocity of around 40\,\kms\
on one epoch (June 1997), although they found a constant 18--19\,\kms\
at several other epochs. They argue that this anomaly is real, perhaps
due to stellar pulsations or a companion in a highly elliptical
orbit. I therefore checked the radial velocity of Gl406 against two
other standards observed, Gl402 and Gl876.  These give
18.8\,$\pm$\,0.1\,\kms\ and 18.2\,$\pm$\,0.5\,\kms\ respectively
(adopting \vrad\,=\,-1.04\,$\pm$\,0.03\,\kms\ for Gl402 from Nidever
et al.\ \cite{nidever02}, and \vrad\,=\,-2.0\,$\pm$\,0.5\,\kms\ for
Gl876 from Delfosse et al.\ \cite{delfosse98}). This is slightly lower
than the adopted value for Gl406. Although the difference is
significant compared to the random errors, it is at a level where
systematic errors could
dominate. Moreover, it has no statistically significant impact on the
derived radial velocity for the secondary template 2M1439+1929: cross
correlation against the Gl402 and Gl876 spectra yields
-26.6\,$\pm$\,1.3\,\kms\ and -27.3\,$\pm$\,1.2\,\kms\ respectively,
compared to the adopted value of -26.3\,$\pm$\,0.50\,\kms.

\section{Rotational broadening via cross correlation}\label{xcor}

UCD rotation velocities are determined via a cross correlation of the
UCD dwarf spectra (the `targets') against spectra of stars with
essentially zero rotation velocity (the `templates'). Under
the assumption that the line broadening of the UCD is dominated by the
rotational broadening (and that the line broadening of the template is
comparatively small), the width of the cross correlation function is a
measure of the \vsini\ of the UCD, which may then be calibrated as
discussed below.  This follows the method used by Tinney \& Reid
(\cite{tinney98}), Mohanty \& Basri (\cite{mohanty03}) and White \&
Basri (\cite{white03}).  The cross correlation is done on each order
separately and the results combined to obtain the final \vsini\
measure.

The two primary templates are Gl402, an M4V with \vsini$<$2.3\,\kms,
and Gl406, an M5.5V with \vsini$<$2.9\,\kms\ (Delfosse et al.\
\cite{delfosse98}).  The procedure implicitly assumes that the UCDs
have similar spectra to the templates, ideally differing only in
\vsini.  This is of course not the case, and the
differences between mid M dwarfs and early/mid L dwarfs could bias
the results.  But in the absence of appropriate L
dwarfs with independently determined rotation velocities this is the
best we can presently do with the cross correlation method. (An
alternative method is to examine individual lines according to a
physical model of line formation and achieve a best fit against a grid
of model spectra, as done, for example, by Schweitzer et al.\
\cite{schweitzer01}, although this was not very sensitive to \vsini).
In some cases, poor fits to the cross correlation peak occur, and such
fits are rejected.  The impact of spectral type mismatch is assessed via
a `bootstrap' method, namely by using the slowest rotating L dwarf
as a secondary template (see
section~\ref{projvsini}).

The cross correlation is done again with {\sc fxcor}, but
independently of the radial velocity determination.  The main peak of
the cross correlation function is fit with a Gaussian and its FWHM
measured.  If a good fit cannot be obtained, or if the peak is
ambiguous, then the order is not used.  

The main uncertainty in performing this fit is determining the
`background' for fitting the Gaussian (i.e.\ the effective zero level
in the cross correlation function). For this reason, three Gaussians
were fit corresponding to the best fit, the minimum FWHM fit and the
maximum FWHM fit.  The fits were performed interactively, with the
maximum and minimum representing conservative limits. The range
between these comprises a generous assessment of errors in the fitting
procedure.  Although the peaks sometimes deviated from a Gaussian
shape (especially at larger rotation velocities), a Gaussian was
nonetheless felt to be the best overall parametrized form among those
tested.

These FWHM values are converted to \vsini\ values through the
following calibration process.  The template spectral orders are
artificially spun up by convolving their spectra with the rotation
profile given by Gray (\cite{gray92}, eqn.\ 17.12) with \vsini\
spanning 10 to 100\,\kms\ in steps of 5\,\kms.  A limb darkening
parameter of $\epsilon=0.6$ is assumed in this profile. The spun up
spectra are cross correlated with the non-rotating template and the
FWHM of the peak is measured in the same way as described above. (Here
the difference between the FWHM of the minimum and maximum Gaussian
fits is much smaller than in the UCD case, so is neglected.)  A linear
least squares fit is then obtained between the FWHM and \vsini,
separately for each order. It was found that the Gaussian was a poor
fit to this cross correlation function above \vsini\,=\,70\,\kms, so
the fit is only made in the range 10--70\,\kms. For a few orders a
good calibration could not be obtained so these were not used.

When this calibration is applied to determine \vsini\ for each order and these
combined (next section), we must pay attention to
the minimum value of \vsini\ to which this procedure is
sensitive. Following Tonry \& Davis (\cite{tonry79}), the measured width of the
cross correlation function is made up of a number of components
\begin{equation}
\sigma^2_{\rm meas} = \sigma^2_{\rm rot} + \sigma^2_{\rm nat} +
2\sigma^2_{\rm inst}
\end{equation}
where $\sigma_{\rm rot}$ is the rotational width, $\sigma_{\rm nat}$
is the natural, or intrinsic, line width in the UCD, and $\sigma_{\rm
inst}$ is the instrumental broadening (two contributions as two
spectra form the cross correlation function). I assume that the
rotational and intrinsic contributions from the template are
negligible. Because the terms are additive, then with `perfect' data,
any non-zero rotational broadening could be detected even with
non-zero $\sigma_{\rm inst}$. In reality, however, small broadenings
cannot be detected due to noise and template/target mismatch. I make
the somewhat ad hoc assumption that we can only detect $\sigma_{\rm
rot}$ if it exceeds the other broadening contributions. If we further
assume that the UCD natural broadening is negligible (reasonable if
resonance lines and gravity sensitive lines are avoided), then the
minimum detectable value of $\sigma_{\rm rot}$ is equal to
$\sqrt{2}\sigma_{\rm inst}$.  The FWHM of the instrumental broadening
is set by the slit width of 7.8\,\kms\ (equal to $1''$; when the
seeing was below this -- as was often the case -- the instrumental
broadening is reduced). Note that the {\it full} width of a rotational
profile corresponds to {\it twice} the rotational velocity (one half
of the line is created by the blueshifted approaching limb of the
star, the other half by the redshifted receding limb). Thus the
minimum detectable \vsini\ is therefore
$\sqrt{2}\,\times\,7.8/2$\,=\,5.5\,\kms.  Any derived values of
\vsini\ below this limit are dropped from the calculations in the next
section. If this limit has been overestimated, then we would
potentially drop too many measures, overestimating the combined
\vsini. (It turns out that I only drop one value for being below this
limit, so this is not significant.)  If this limit has been
underestimated, then additional values may need to be dropped,
potentially raising \vsini.

\section{Projected rotational velocities}\label{projvsini}

The calibrations described in the previous section are applied to each
order to determine a \vsini\ for each of the three fits.  The
different orders are then combined to give the mean, minimum and
maximum values as listed in Table~\ref{vsini_table}.  The mean value
is a weighted mean of the `best fit' \vsini\ values described in the
previous section, using a weight of 1,2,3 or 4 depending on the
quality of the fit to its cross correlation peak. Obvious outliers are
clipped when forming this mean (7\% of all orders/spectra are clipped
in this way).  The minimum and maximum values listed in
Table~\ref{vsini_table} are likewise a weighted mean of the minimum
and maximum derived \vsini\ values.

\begin{table*}
\begin{center}
\parbox{\textwidth}{
\caption{\vsini\ values for ultra cool dwarfs observed in this survey.
The first column lists the full name of each target: an abbreviated
version is used elsewhere in this paper, where 2M=2MASS, DE=DENIS and
SD=SDSS.  The second and third columns list the spectral types and I
magnitudes taken from the literature. (I magnitudes given to only one
decimal place have been estimated.)  Columns 4--7 and 8--11 list the
values of \vsini\ deduced via cross correlation with the M dwarf
templates and L dwarf template (2M1439+1929) respectively.  ``mean''
(i.e.\ \mvs) is a weighted mean formed by combining measurements from
individual spectrograph orders and/or exposures: $N$ orders are
retained in the final calculation. ``min.'' and ``max.'' are the
minimum and maximum values of \vsini\ similarly averaged over all
orders, and so encompass generous estimates of the uncertainty in
fitting the peak of the cross correlation function.  The final column
gives the radial velocities relative to the solar system
barycentre. In addition to the random errors given for these, there
may be an additional systematic error from the zero point uncertainty
(due to possible slit decentering) of no more than 1\,\kms.  A colon,
``:'', indicates an uncertain value. }
\label{vsini_table}
}
\begin{tabular}{llrlrrrrrrrrr@{.}l}
\hline
Name      & &    SpT &   $I$ &   \multicolumn{8}{c}{\vsini\ / \kms} & \multicolumn{2}{c}{\vrad\ / \kms} \\ \cline{5-12}
          & &        &       &   \multicolumn{4}{c|}{M dwarf template}  &  \multicolumn{4}{|c}{2M1439+1929 template} \\
	  & &	    &       &	mean  &	min. & 	max. &	$N$  &	\multicolumn{1}{|l}{mean} &	min. &  max. &	$N$  \\
\hline
DENIS-P & J1431-1953&           M9.0&	17.47&	47.4&   44.1&	52.6&	 1&	37.1&	33.7&	41.6&	 4  & -11&4\,$\pm$ 2.8\\
DENIS-P & J1159+0057&	        L0.0&	17.32&	35.2&	31.9&	39.7&	 3&	74.5&	69.1&	83.5&	 1  &  -3&0\,$\pm$ 1.5 :\\
2MASSI & J0746425+200032&	L0.5&	15.11&	27.3&	25.6&	30.6&	 6&	25.8&	23.0&	28.8&	 9  &  54&1\,$\pm$ 0.9\\
2MASSW & J1412244+163312&	L0.5&	17.1&	17.3&	14.7&	19.2&	 6&	16.4&	13.9&	19.2&	 4  &   6&9\,$\pm$ 0.8\\
2MASSW & J1439284+192915&	L1.0&	16.12&	11.2&	9.6&	12.8&	 9&	--&	--&	--&	--  & -26&3\,$\pm$ 0.5\\
DENIS-P & J1441-0945&	        L1.0&	17.32&	17.4&	14.8&	20.6&	 4&	15.9&	13.3&	19.3&	 6  & -27&9\,$\pm$ 1.2\\
2MASSW & J1145572+231730&	L1.5&	18.62&	12.5&	10.5&	14.4&	 9&	12.7&	10.5&	14.8&	 6  &   3&7\,$\pm$ 0.9\\
2MASSW & J1334062+194034&	L1.5&	18.76&	25.2&	22.1&	28.1&	 4&	25.4&	21.5&	29.9&	 5  &  -4&3\,$\pm$ 2.1 :\\
2MASSI & J1029216+162652&	L2.5&	17.9&	28.3&	23.6&	32.5&	 7&	28.0&	22.8&	33.1&	 8  & -29&2\,$\pm$ 4.0\\
DENIS-P & J1047-1815&	        L2.5&	17.75&	16.3&	15.0&	18.3&	12&	15.0&	12.2&	17.6&	11  &   6&0\,$\pm$ 0.8\\
2MASSW & J0913032+184150&	L3.0&	19.07&	15.0&	11.9&	17.6&	11&	20.3&	18.0&	22.6&	 3  &  28&4\,$\pm$ 2.5 :\\
SDSSp  & J120358.19+001550.33&	L3.0&	18.88&	27.6&	24.1&	32.3&	 7&	31.7&	25.6&	36.8&	 8  &  -2&7\,$\pm$ 2.3\\
2MASSW & J1615441+355900&	L3.0&	18.1&	12.1&	10.7&	13.8&	 5&	12.8&	 9.4&	16.0&	10  & -20&2\,$\pm$ 0.9\\
2MASSW & J2224438-015852&	L4.5&	18.02&	25.4&	22.2&	27.9&	 3&	24.7&	20.9&	29.2&	 6  & -37&4\,$\pm$ 3.4\\
2MASSW & J1507476-162738&	L5.0&	16.65&	27.1&	22.6&	31.6&	 4&	27.2&	23.4&	32.7&	 4  & -39&3\,$\pm$ 1.5\\
2MASSI & J0825196+211552&	L7.5&	19.22&	--&	--&	--&	--&	16.9&	11.3&	21.4&	 6  &  20&5\,$\pm$ 2.0\\
\hline
\end{tabular}
\end{center}
\end{table*}

As discussed in the previous section, the cross correlation method
implicitly assumes that the template and target spectra differ only in
the rotation velocity, yet the templates are mid M dwarfs whereas the
targets range in spectral type from M9 to L5 with one L7.5. This
potentially biases the \vsini\ determinations. To
partially overcome this, the entire cross correlation and calibration
procedure is repeated using the slowest rotating L dwarf as a
template. The relevant star is 2M1439+1929, which was also used as a
template by Basri et al.\ (\cite{basri00}) for the same reason and purpose.
Fortunately, this L1.0 dwarf is bright enough to use as a
template and is a reasonably early L type such that its cross
correlation against the mid M dwarfs was probably not too erroneous.
Of course, with \mvs=11.2\,\kms, this L dwarf is rotating faster than
the M dwarf templates.  Using a rotating template will generally
underestimate \vsini, because when we spin up
the templates for the calibration, a given \vsini\ corresponds to a
higher FWHM than would be the case with a non-rotating template.  The
derived \vsini\ values using 2M1439+1929 as the template are shown in
Table~\ref{vsini_table}. Other than in one case (and that is a single
measure of lowest quality), there is actually no significant tendency
for \vsini\ to be lower (or higher) than the
values derived with the M dwarf templates.
Indeed, the agreement between the values is good, deviating by less
than 2\,\kms\ in 10 out of 14 cases. In two cases (DE1431-1953 and
DE1159+0057) the disagreement is more than 10\,\kms, but in both of
these cases one of the measurements is based on only a single (poor
quality) measurement. In the remaining two cases (2M0913+1841 and
SD1203+0015) the discrepancy is around 5\,\kms\ but the minimum to
maximum ranges just about overlap (and both stars are among the
faintest observed).  Thus although the \vsini\ calibration using
2M1439+1929 has used the M dwarf calibration to identify this
template, the spectra and fitting procedures are independent.
The fact that both templates give consistent \vsini\
ranges indicates that the impact of template mismatch it not major, at
least not when using multiple orders across a wide wavelength
range.\footnote{This consistency could be misleading if use of the L
template gives a positive bias to the derived \vsini\ (compared to the
M template) which is then offset by a negative bias from having used a
{\it rotating} L template. However, it is not obvious that the
later-type template should give a {\it positive} bias and, moreover,
the two biases would have to contrive to have the same magnitude.}  Fig.~\ref{specplot} further shows that there is relatively
little mismatch between even an M5.5V and L3.0V. Only three of the
UCDs are later than this. For the latest type (L7.5V), no acceptable
cross correlation function could be obtained against the M5.5V,
perhaps indicative of commencing mismatch problems.  In
the rest of this article I adopt the 2M1439+1929 cross correlation
values, except of course for 2M1439+1929 itself and except for
DE1159+0057 because it has only a single poor quality measurement
against this template.

The most striking result from Table~\ref{vsini_table} is that L dwarfs
are relatively fast rotators, considerably faster than is typically
found for M dwarfs.  They all have \mvs\ in the range 11--37\,\kms.
Moreover, the lowest value of \mvs\ is 11.2\,\kms, and the lowest
minimum value is 9.4\,\kms, well above the minimum detection limit of
5.5\,\kms\ (section~\ref{xcor}). 
This agrees with the results of Mohanty \& Basri (\cite{mohanty03}), who
found no L dwarf in their sample of 13 to have \vsini\ below 10\,\kms.
Two specific targets, 2M0746+2000 and 2M1439+1929, have
\vsini\ determined by others: The former has $24\pm5$\,\kms\ according
to Reid et al.\ (\cite{reid02}) and $20\pm10$\,\kms\ according to
Schweitzer et al.  The latter is derived to have $10\pm2.5$\,\kms\ by
Basri et al.\ (\cite{basri00}) and $10\pm2$\,\kms\ by Mohanty \& Basri
(\cite{mohanty03}). These agree well with the values and ranges
derived in the present work.

\begin{figure}
\centerline{
\psfig{figure=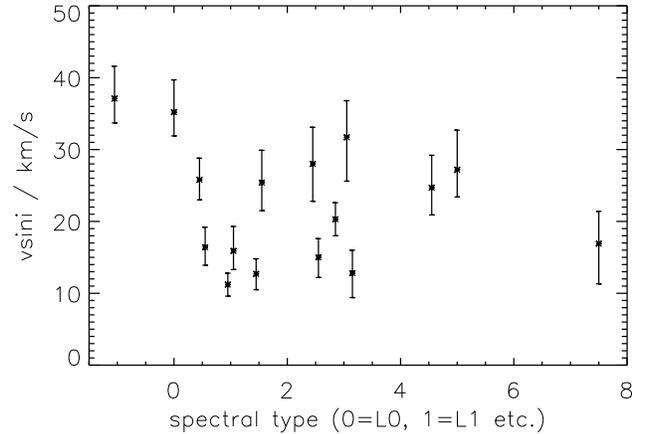,angle=90,width=0.50\textwidth}
}
\caption{Projected rotational velocities for ultra cool dwarfs
observed in this survey. The values are for the cross correlation
against the L dwarf template, 2M1439+1929 (columns 8, 9 and 10 of
Table~\ref{vsini_table}) except for 2M1439+1929 itself and
DE1159+0057, for which results against the M dwarf templates are
shown.  The cross symbol shows the expected value and the `error bars'
are the maximum and minimum values. Objects with equal spectral types are
slightly offset on the spectral type axis for clarity.}
\label{vsini_fig}
\end{figure}

The projected rotational velocities show no trend with spectral
type (Fig.~\ref{vsini_fig}), as is also seen in the results from
Mohanty \& Basri (\cite{mohanty03}) over this spectral type range.  There is
similarly no correlation with I magnitude.

\section{Limits on rotation periods}\label{limits}

The projected rotation velocity, \vsini, is a lower limit to the
equatorial rotation velocity, $v$.  Assuming that a star rotates as a
rigid sphere, $v= 2 \pi R / T$, where $R$ is the equatorial radius of
the star and $T$ is its rotation period.  Given $R$, we can therefore
derive the maximum rotation period, \tmax. More usefully, we can also
derive the expected period and `likely' range of periods consistent
with \vsini. This may then be compared with published periods obtained
from photometric monitoring.

Rigid body rotation is conventionally supported by the theoretical
argument that these cool stars are probably fully convective (e.g.\
Chabrier \& Baraffe \cite{chabrier97}) although counter arguments
exist (Mullan \& MacDonald \cite{mullan01}).  Evolutionary models show
that a few hundred million years after formation, UCDs with masses
between 0.04\,\msol\ and 0.09\,\msol\ all have radii around
0.1\,\rsol\ (Chabrier et al.\ \cite{chabrier97}). Specifically,
between 0.5\,Gyr and 1.0\,Gyr the radii are in the range
0.10-0.12\,\rsol, shrinking to 0.085--0.115\,\rsol\ at 5\,Gyr.  As the
observed field UCDs reported here are probably at least 1\,Gyr old
(Gizis et al.\ \cite{gizis00}), I adopt a radius of 0.1\,\rsol\ for
all targets when deriving period ranges.  The effect of deviations
from this are discussed below, as are limits on $R$ which can be
derived from the data.  In useful units, the relation between \vsini,
$T$ and $R$ is
\begin{equation}
T/{\rm hrs} = 121.47 \, \frac{{\rm R}}{0.1{\rm R}_{\odot}}  \, \frac{1}{v/{\rm km\,s}^{-1}}
\label{period}
\end{equation}

Assuming that the spin axes of L dwarfs are randomly oriented in
three-dimensional space, it can be seen that the probability,
P($i$)\,d$i$, of observing a spin axis between $i$ and $i+{\rm d}i$ is
$\sin i \, {\rm d} i$. From this we may simply derive that $<\sin i> =
\pi/4=0.785$, and hence that $<v>$\,=\,1.27$<v \sin i>$ from which we
calculate $<T>$ from the above equation.  The largest possible value
of $T$, \tmax, occurs when $\sin i = 1.0$, i.e.\ when the star is
viewed edge on. The minimum value of $T$ is arbitrarily small (as in
the limit as $\sin i$ tends toward zero $v$ tends toward
infinity). A more useful statement concerns the probability, $f$, that
$v$ is not above some value, $v_f$.  By fixing $f$ at 0.9 we may then
say that $v$ is less than $v_f$ with 90\% confidence. Generally,
\begin{equation}
v_f = (1-f^2)^{-1/2} \, v \sin i
\end{equation}
so for $f=0.9$, $v_f=2.29\,$\vsini. 
The corresponding 90\% confidence lower limit on the period, \t90,
can then be determined from equation~\ref{period}.

\begin{figure}
\centerline{
\psfig{figure=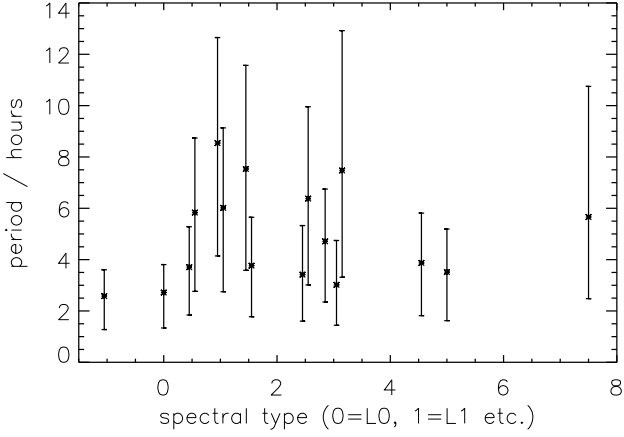,angle=90,width=0.50\textwidth}
}
\caption{Derived period limits for the ultra cool dwarfs from the data
given in Table~\ref{period_table} for the 2M1439+1929 template (except
for 2M1439+1929 and DE1159+0057, which are for the M dwarf
templates). The star symbol is the expected period; the `error bars' show
the maximum period and 90\% confidence `minimum' period.}
\label{period_fig}
\end{figure}

These period limits are calculated from the \vsini\ values in
Table~\ref{vsini_table} to give the period ranges reported in
Table~\ref{period_table} and Fig.~\ref{period_fig}: the maximum
period, \tmax, is calculated from the minimum value of \vsini, and the
minimum period, \t90, is calculated using the maximum value of
\vsini. The expected period, \texp, is calculated from the mean value
of \vsini.  Formally, the range \t90:\tmax\ is the 90\% confidence
interval for the period, and is one-sided because \tmax\ is an
absolute maximum (for fixed radius).  However, \t90\ is a rather
conservative lower limit to the period because the statistical projection
argument has been applied to the upper limit of \vsini.

\begin{table*}
\begin{center}
\parbox{9.5cm}{
\caption{Expected, \texp, maximum, \tmax, and `minimum', \t90,
rotation periods for the ultra cool dwarfs taking into account both
the statistical distribution of $i$ and the range of \vsini\ given in
Table.~\ref{vsini_table}. \texp\ is calculated using the mean \vsini\
and the expected value of $\sin i$ ($=\pi/4)$. \tmax\ is derived from
the minimum \vsini\ and maximum $\sin i$ ($=1$).  \t90\ is calculated
from the maximum \vsini\ and from that value of $i$ (=26\deg) for
which there is a 90\% chance that the inclination lies above this:
i.e., there is only a 10\% chance that the period is shorter than
\t90\ when adopting the maximum \vsini\ (and even less chance when adopting
a more likely \vsini). }
\label{period_table}
}
\begin{tabular}{lrrrrrr}
\hline
 Name        & \multicolumn{6}{c}{rotation period / hrs} \\ \cline{2-7}
             & \multicolumn{3}{c|}{M dwarf template} & \multicolumn{3}{|c}{2M1439+1929 template} \\
             &  \texp &  \t90 & \tmax   &   \multicolumn{1}{|l}{\texp} &  \t90 & \tmax   \\
\hline
\object{DE1431-1953} &  2.02 &  1.01 &  2.75  &   2.58 &  1.28 &  3.60 \\
\object{DE1159+0057} &  2.72 &  1.34 &  3.81  &   1.28 &  0.64 &  1.76 \\
\object{2M0746+2000} &  3.50 &  1.73 &  4.74  &   3.71 &  1.84 &  5.28 \\
\object{2M1412+1633} &  5.53 &  2.76 &  8.26  &   5.83 &  2.76 &  8.74 \\
\object{2M1439+1929} &  8.54 &  4.14 & 12.65  &  -- & -- & -- \\
\object{DE1441-0945} &  5.50 &  2.57 &  8.21  &   6.02 &  2.75 &  9.13 \\
\object{2M1145+2317} &  7.65 &  3.68 & 11.57  &   7.53 &  3.58 & 11.57 \\
\object{2M1334+1940} &  3.80 &  1.89 &  5.50  &   3.77 &  1.77 &  5.65 \\
\object{2M1029+1626} &  3.38 &  1.63 &  5.15  &   3.42 &  1.60 &  5.33 \\
\object{DE1047-1815} &  5.87 &  2.90 &  8.10  &   6.38 &  3.01 &  9.96 \\
\object{2M0913+1841} &  6.38 &  3.01 & 10.21  &   4.71 &  2.35 &  6.75 \\
\object{SD1203+0015} &  3.47 &  1.64 &  5.04  &   3.02 &  1.44 &  4.74 \\
\object{2M1615+3559} &  7.90 &  3.84 & 11.35  &   7.47 &  3.32 & 12.92 \\
\object{2M2224-0158} &  3.77 &  1.90 &  5.47  &   3.87 &  1.82 &  5.81 \\
\object{2M1507-1627} &  3.53 &  1.68 &  5.37  &   3.52 &  1.62 &  5.19 \\
\object{2M0825+2115} & -- & -- & --  &   5.66 &  2.48 & 10.75 \\
\hline
\end{tabular}
\end{center}
\end{table*}

\section{Assessment of published variability and period data: individual objects}
\label{assessment}

Of the 16 UCDs with \vsini\ measured in this work, nine have
variability or period data published by one or more authors:
Bailer-Jones \& Mundt (\cite{bjm01}, hereafter BJM); Bailer-Jones \&
Lamm (\cite{bjl}); Clarke et al.\ (\cite{clarke02a}; \cite{clarke02b},
hereafter COT); Gelino et al.\ (\cite{gelinoetal02}, hereafter G02);
Gelino (\cite{gelino02}).  A further five UCDs have variability/period
data for which \vsini\ values have been published by others.  The data
for these 14 UCDs are now compared.  All variability data is in an I
band unless stated otherwise.

{\bf 2M0746+2000}.~~This was identified by Reid et al.\ (\cite{reid01}) as a
near equal mass binary with a separation of 0.22$''$ (2.7\,AU) through
HST imaging, with a magnitude difference of $\Delta$I=0.6.  There is no
evidence for binarity from my spectral data which is presumably
dominated by the brighter component.  COT identify significant
variability in their light curve data with an amplitude of 0.007 mag
and derive a period of about 3\,hrs, although they underline that this
interpretation may be complicated by the binarity.  It
is also based on just a single dimming in the data (the source was
only monitored for 6.5\,hrs).  Nonetheless, this period is consistent
with the period limits in Table~\ref{period_table}.  G02 report a
period of 31\,hrs for this source with a confidence of almost
5$\sigma$.  This is inconsistent with the much smaller maximum period
of 5.7\,hrs in Table~\ref{period_table}.  This maximum is based on a
cross correlation peak clearly widened above the the minimum \vsini\
threshold and there is good consistency between the individual orders
and the two templates.  The 31 hour period of G02 can therefore be
ruled out as the rotation period. The rotation period and \vsini\
could only be made consistent if the star had a radius of at least
0.6\,\rsol.  This would require the star to be either much more
massive than an L dwarf or very young, perhaps only a few million
years from its birth line (Chabrier \& Baraffe \cite{chabrier97}; see also Joergens
et al.\ \cite{joergens03}). However, in a detailed study based on kinematics,
spectroscopy and photometry, Reid et al.\ (\cite{reid00}) conclude that
2M0746+2000 has a mass between 0.07\,\msol\ and 0.09\,\msol\ and an age of at
least 1\,Gyr.  A probable explanation for the 31 hour period from G02
is that it is not a real period (a 5$\sigma$ detection\ is not large
for a periodogram).  Alternatively, it could be an alias of a shorter
period. A third possibility is that this period is not related to the
rotation at all, and may not even be a stable period.  It could
instead be an artifact of unstable light curve modulation due to
non-static surface features, as was suggested for some UCDs in
BJM. Gelino (\cite{gelino02}) also reports J,H,K monitoring observations of this
star. Although there is formally no variability across the full
light curves, there is strong evidence for a rise of 0.02 mag in J and
K lasting 1.5\,hrs (over a total time span of 5.5\,hrs: see Fig.~4.5
in that work).\footnote{Gelino states that this is not significant,
but this conclusion appears to be based on comparing the amplitude
with the photometric error in a {\it single} measurement.  Yet the
1.5\,hrs rise is observed in about 30 points systematically displaced
from the mean, and is clearly a significant deviation.} Interestingly,
this rise is similar in duration to the dip seen in the same object by
COT.

{\bf 2M1412+1633}.~~G02 report a non-detection of variability on
scales above 0.025 mag (1.4 times the reported RMS) on time scales of
up to 85 days (although most data were obtained over 7 days according
to Fig.~3.7 of Gelino \cite{gelino02}).  Likely periods from
Table~\ref{period_table} could in principle have been detected,
although the time sampling of G02 is very sparse so
sensitivity to the likely periods is low.  If there were sufficient
temporal sensitivity, then any net rotational modulation must have
been smaller than 0.025 mag.

{\bf 2M1439+1929}.~~This was not detected as variable by any of three
independent monitoring programs: BJM derive an upper limit of 0.01 mag
over timescales of 1\,hrs to 100\,hrs; a slightly higher upper limit over
longer timescales comes from G02. Bailer-Jones \& Lamm (\cite{bjl}) restrict
the J and K band variability to be less than 0.04 mags on time scales
between 20 min and 13 days. All of these programs had sampling dense
enough to detect likely rotation periods from
Table~\ref{period_table}. I conclude that the amplitude of rotational
modulation is less than 0.01 mag in I and 0.04 mag in J and K.

{\bf 2M1145+2317}.~~While BJM reported a period of
11.2\,$\pm$\,0.8\,hrs, a second monitoring epoch reported in the same
paper confirmed the variability but not the period. This lack of
stability led BJM to suggest that the modulating pattern on the star
was not stable: if surface features were evolving during the second
epoch on a timescale less than the rotation period, this could have
`masked' the rotation period. The present \vsini\ data suggest a
period between 3.6\,hrs and 11.6\,hrs, consistent with the 11.2\,hrs period
and this masking hypothesis.  (It is not possible that the modulating
features were simply weaker in the second epoch as then no variability
would have been found.)  A period of 11.2\,$\pm$\,0.8\,hrs corresponds
to an equatorial rotation speed of 10.8\,$\pm$\,0.8\,\kms\
(eqn.~\ref{period}).  Combining this with the \vsini\ range of
10.5--14.8\,\kms\ implies that the inclination angle, $i$, lies in the
range 62\deg--90\deg, with this lower limit being rather
conservative. The expected value is $i$=76\deg.
Thus if the rotation period of 11.2\,$\pm$\,0.8\,hrs from BJM is
correct, 2M1145+2317 would appear to be a near edge-on rotator (which
makes detection of rotation by monitoring more likely). If the stellar
radius were larger, this would decrease the expected and minimum
inclination angles (and eventually the maximum too), and vice versa.
Alternatively, we can use this period plus the minimum value of
\vsini\ (10.5\,\kms) to place a lower limit on the radius (i.e.\
corresponding to $i$=90\deg) of this L1.5 dwarf. This gives $R_{\rm
min}=0.097 \pm 0.007$\,\rsol, the uncertainty arising from the
uncertainty in the period. This is consistent with structure models for
a large range of ages (see section~\ref{limits}).

{\bf 2M1334+1940}.~~The variability detection of this object was one
of the most significant in BJM; the detected period of
2.68\,$\pm$\,0.13\,hrs was relatively significant (12$\sigma$).  This
is consistent with the present period deductions and corresponds to an
equatorial rotation speed of 45.3\,$\pm$\,2.2\,\kms.  Combining this
with the \vsini\ range from Table~\ref{vsini_table} of
21.5--29.9\,\kms\ implies a range of $i$=27\deg--44\deg, with
$i$=34\deg\ the expected value.  Although this star is quite faint, we
can have some confidence in this given the agreement in \vsini\ values
for the two templates and the reasonable significance of the period
detection. As in the previous case, we can instead use the data to
derive the minimum radius, which is $R_{\rm min}=0.047 \pm
0.002$\,\rsol. This is a smaller limit than in the previous case,
because these data permit smaller inclination angles for 2M1334+1940.

{\bf 2M1029+1626}.~~G02 report no variability in this L2.5 dwarf,
although as this is based on only a few observations per night
(Fig.~3.4 of Gelino \cite{gelino02}), this program would probably not
have been sensitive to the short periods deduced in
Table~\ref{period_table}.

{\bf 2M0913+1841}.~~This was found to be variable by BJM but with no
significant period.  These observations should have been sensitive to
the likely periods reported here (although plausibly the sampling in
BJM could have thwarted its detection: see Fig.~1 in that article).
Assuming that this variability is due to changes in photospheric
features, then these features must be changing in pattern and/or
brightness at least as fast as the likely rotation periods
(3--10\,hrs) in order to mask a rotationally modulated signature.
This is the conclusion of BJM based on the {\it assumption} that this L
dwarf had a rotation period in the range of detectability. That this is
now demonstrated to be likely adds some strength to that conclusion.

{\bf SD1203+0015}.~~As with 2M0913+1841, this was found to be a
non-periodic variable by BJM. On the other hand, it was intensively
monitored by Bailer-Jones \& Lamm (\cite{bjl}) in the J and K bands, who
found no variability above the larger limit of 0.04 mags over
timescales of 20\,min to 7\,days. The same comments as for 2M0913+1841
therefore apply.

{\bf 2M2224-0158}.~~G02 report no variability, but as with
2M1029+1626, there were only a few observations over each of a few
nights, so periods of a few hours could have remained undetected.

There are five other UCDs with published \vsini\ values which also
have variability data in the literature. These are now briefly
discussed to complete this survey of UCD rotation and variability data.

{\bf \object{2M1146+2230}~(L3V)}.~~This is an optical binary with a
separation of 0.29$''$ (7.6\,AU) and $\Delta$I=0.3\,mag (Reid et al.\
\cite{reid01}).  Mohanty \& Basri (\cite{mohanty03}) determine
\vsini\,=\,32.5\,$\pm$\,0.2\,\kms, presumably for the brighter
component.  BJM detected marginally significant variability, with
possible rotation periods at 5.1\,$\pm$\,0.1\,hrs (15$\sigma$) and
3.0\,$\pm$\,0.1\,hrs (6$\sigma$). The former is slightly too long for
the measured \vsini, although it would be consistent if the radius
were 0.13\rsol\ (more if the star is not observed edge on). This
radius is consonant with the evolutionary models, especially if this L
dwarf is less than 1\,Gyr in age (Chabrier \& Baraffe
\cite{chabrier97}).  The latter period would also be consistent, but
is of too marginal significance to warrant further discussion.
Neither G02 nor COT detected variability in this source, although the
upper limits of both are consistent with the BJM detection.  Given the
sampling, either BJM or COT could have detected the likely periods,
but the weight of evidence is that the modulation contrast was not
large enough (or not stable enough) to permit detection.

{\bf \object{LP944-20}~(M9V)}.~~\vsini\ was determined to be
28.3\,$\pm$\,2\,\kms\ by Tinney \& Reid (\cite{tinney98}) and
39\,$\pm$\,2\,\kms\ by Mohanty \& Basri (\cite{mohanty03}). Tinney \&
Tolley (\cite{tinney99}) detected variability with an amplitude of
0.04 mag and timescale of 2\,hrs.  (With a p value of only 0.02 this
would not have been counted as variable on the criteria of BJM or G02,
although the fact that the detection is made in two narrow bands
compensates for this.)  Although these authors did not associate this
with a rotation period, it is consistent with the maximum rotation
periods of 4.3\,hrs and 3.1\,hrs implied by the above \vsini\ values.

{\bf \object{DE1228-1547}~(L4.5V)}.~~Tinney \& Tolley
(\cite{tinney99}) did not detect variability in this, although their
observations were in principle sensitive to the periods commensurable
with the \vsini\ of 22\,\kms\ reported by Basri et al.\
(\cite{basri00}) and Mohanty \& Basri (\cite{mohanty03}). This implies
low contrast surface features.

{\bf \object{BRI0021-0214}~(M9.5V)}.~~Tinney \& Reid (\cite{tinney98})
established \vsini\,=\,42\,$\pm$\,8\,\kms\ for this compared to
34\,$\pm$\,2\,\kms\ from Mohanty \& Basri (\cite{mohanty03}).  Martin
et al.\ (\cite{martin01}) detect several periods in this source from I
band monitoring, in particular at 4.8\,hrs and 20\,hrs. Both are too
long for the measured \vsini\ range unless the radius is at least
0.13\,\rsol\ and 0.5\,\rsol\ respectively -- more if the star is not
observed edge on -- something which Martin et al.\ rule out. This is
taken as evidence for the masking hypothesis.

{\bf \object{Kelu-1~(L2V)}}.~~This is a very rapid rotator, as reported both by
Basri et al.\ (\cite{basri00}; \vsini\,=\,60\,$\pm$\,5\,\kms) and
Mohanty \& Basri (\cite{mohanty03}; \vsini\,=\,60\,$\pm$\,2\,\kms).
Clarke et al.\ (\cite{clarke02a}) detected Kelu-1 to be periodically
variable with a period of 1.80\,$\pm$\,0.05\,hrs and peak-to-peak
magnitude of 0.011 mag.  This is consistent with the \vsini\ values,
and is discussed at length by Clarke et al.


\section{Conclusions}
I have presented \vsini\ measurements for 16 ultra cool field dwarfs
with spectral types between M9 and L7.5. 14 of these are new,
doubling the number of L dwarfs with measured \vsini.  14 UCDs have
\vsini\ between 10 and 30\,\kms, the other two between 30 and
40\,\kms. This confirms a previously published result that, compared
to the M dwarfs, L dwarfs are fast rotators, with implications for their
angular momentum evolution.  In addition to
establishing mean \vsini\ values I determined a conservative \vsini\
range for each UCD which encompasses the asymmetric uncertainties in
fitting the cross correlation peak: this full range is typically
7\,\kms.

With theoretically predicted radii of 0.1\,\rsol, the expected
rotation periods of these UCDs (i.e.\ using $<\sin i> =\pi/4$) is
3--10\,hrs.  The measured \vsini\ determines the maximum rotation
period. Using the statistical distribution of $i$, a one-sided 90\%
confidence range on the period of each UCD can be established: this is
5.3--12.1\,hrs and 1.8--4.0\,hrs for \vsini=10\,\kms\ and 30\,\kms\
respectively. A larger radius for a given \vsini\ would imply longer
periods.

From this analysis, suggested periods for UCDs from published
monitoring programs were assessed. The following conclusions were
drawn.  The period of 11.2\,$\pm$\,0.8\,hrs for 2M1145+2317 detected
by Bailer-Jones \& Mundt (\cite{bjm01}) is likely to be the rotation
period for this L1.5 dwarf, and the inclination angle is constrained
to be $i$=62\deg--90\deg, i.e.\ a near edge-on rotator. Likewise, the
period of 2.7\,$\pm$\,0.1\,hrs detected for the L1.5 dwarf 2M1334+1940
by the same authors is confirmed as the likely rotation period, and
the inclination is $i$=27\deg--44\deg. The period of 31\,hrs detected
by Gelino et al.\ (\cite{gelino02}) for the brighter component of the
binary 2M0746+2000 can be strongly ruled out as a rotation period.  In
three cases (2M0913+1841, 2M1145+2317 and SD1203+0015), variability
but no period was detected by monitoring programs, even though the
present work derives likely periods which should have been detectable
by the programs. Although all of these objects are faint, this seems
to support the `masking hypothesis', the idea that surface features
evolving in distribution and/or brightness faster than the rotation
period could mask such a period from being detected by monitoring. The
challenge to ongoing work remains to characterise this evolution.

\section*{Acknowledgements}

I would like to thank Andreas Kaufer for assistance during the
observations and for advice regarding the reduction of UVES data.
The constructive comments from the referee are gratefully
acknowledged.


\end{document}